\documentclass[preprint,aps,groupedaddress,amsmath,amssymb,superscriptaddress,footninbib]{revtex4}

\usepackage{graphicx}

\newcommand{\ev}[1]{\left\langle #1 \right\rangle}

\newcommand{\ket}[1]{\mid #1 \rangle}

\begin{document}

\title{Condensation of area quanta ensembles with quantum statistics in Schwarzschild spacetimes}

\author{Ryley McGovern, Seth Major, Trevor Scheuing, Thomas Takis}
\affiliation{Department of Physics, Hamilton College, Clinton NY
13323 USA}

\date{January 2025}

\begin{abstract}

Near-horizon (equivalently high acceleration) observers in spherically symmetric black hole spacetimes have a particularly simple form of the quasilocal energy. Using this energy and indistinguishable area quanta satisfying quantum statistics, a statistical mechanical description of the Schwarzschild black hole geometry for uniformly accelerating observers is developed.  The resulting model has several phases including one with highly excited states, Bose-Einstein condensates, condensates distinct from the usual Bose gas, and degenerate Fermi gases.  In the large area limit, relevant for comparison to the Bekenstein-Hawking entropy, the new condensed state is favored over Bose-Einstein condensation and the degenerate Fermi gas. The entropies of the phases, and the entropy of mixing, are computed. The resulting low-entropic condensed state, in which the quanta are essentially all in the lowest Bose energy state, provides the framework for the quantization of near-horizon geometric fluctuations, which is explored in \cite{MRT}. 
    
\end{abstract}

\maketitle

\section{Introduction}

While classically spherically symmetric black holes are completely characterized by one parameter - the asymptotic mass $M$, the semiclassical environment is significantly more subtle: Hawking showed that local static observers in these spacetimes are bathed in a thermal bath of temperature $T_{loc} = (8 \pi M)^{-1} (1- 2M/r)^{-1/2}$. Bekenstein argued that the second law of thermodynamics is preserved if the black holes acquire entropy proportional to the horizon area, $A$ \cite{bekenstein}. The resulting Bekenstein-Hawking entropy $S_{BH} = A/4 \hbar$ is a key theoretical window into quantum gravity.  (We adopt units where $k=G=c=1$.)

Many developing quantum gravity theories have successful calculations of the Bekenstein-Hawking entropy.  The explanations are diverse and without clear agreement on the microscopic constituents. For example due to the causal cloaking of the horizon and vacuum fluctuations of field theory, area-proportional entropy accumulates via entanglement \cite{entangleentropysorkin,BKLS}.  Another derivation in the context string theory, shows that a calculation of the entropy of a gas of D-branes gives the Bekenstein-Hawking entropy for five-dimensional extremal black holes -  including the factor of 1/4 - due to the protection of the special (BPS) states \cite{SV}: The state counting is preserved as the gravitational coupling is increased and the D-brane ensemble collapses into a black hole configuration.  Loop quantum gravity (LQG) calculations use quanta of area as the microscopic constituents with calculations giving the correct scaling of the black hole entropy with the horizon area.  Typically in these LQG calculations a free parameter of the theory, the Barbero-Immirzi parameter, is fixed to an $O(1)$ value to match the $A/4 \hbar$ result of Bekenstein and Hawking. While there is a remarkable consistency in the proportionality of entropy to horizon area, the calculations differ strongly on what is counted. We lack even a rudimentary understanding of the complementary of these successful approaches.  While the entropy is clearly associated to the spacetime geometry, the physical fundamental constituents of geometry remain mysterious.

In this paper we begin the process of building a statistical model of the Bekenstein-Hawking entropy starting from indistinguishable quanta of geometry and fluctuations in the near horizon geometry \footnote{An observable signature associated to these fluctuations is explored in \cite{MRT}.}. We begin using the discrete area spectrum of LQG and building a model of the statistics of geometric quanta.  Although we follow earlier studies \cite{GNP,asinetal} in their operational approach of a family of observers just outside the horizon and area quanta, our approach differs from this work in two ways: First, we assume a `projection' or garden variety angular momentum degeneracy rather than the `holographic' degeneracy, where the degeneracy scales as $\exp(A/4 \hbar)$. Second, we assume that the particle statistics is determined by spin.  

Particles statistics arises from the fundamental nature of the constituents.  In a room of gas one may be able to locate certain molecules through fluctuations but the correct statistical distribution is given by counting the states of the fundamentally indistinguishable molecules with entropy given by the Sackur-Tetrode relation for an ideal gas. If instead one assumes the gas particles are distinguishable, the entropy would increase simply by adding a partition. This sort of Gibbs paradox does not exist for indistinguishable particles. We adopt this fundamental assumption for the quanta of geometry.  Without a complete theory of quantum gravity we cannot show that the quanta are indistinguishable. Rather it is an assumption that we view as a potentially interesting avenue to explore. 

The LQG approach to black hole entropy was initiated in a series of papers by Krasnov \cite{entropykirill,entropykirill2} and Rovelli \cite{entropycarlo,entropycarlo2}.  The widely accepted approach to the black hole entropy in LQG was formulated by Ashtekar, Baez, Corichi, and Krasnov \cite{ABCK,ABK} (See the introduction of \cite{agulloetal} for a review of the historical development of the subject and \cite{alejandro_rev} for a full review.) This early work was followed by development  in many directions including, and most relevantly to this work, particle statistics in Refs. \cite{GNP,asinetal}. 

In many of the LQG papers on black hole entropy the atoms of geometry are distinguishable. This follows, for instance, from the fact that the atoms are connected via the graph of the spin network state to the exterior geometry and thus might be localizable by suitable observations in the exterior \cite{entropycarlo}.  Or, ``In quantum geometry, each intersecting edge determines the `local' shape of the surface.  Not only does a label [$j$] determine the area contribution of a single edge, but the vertex also can determine the
local curvature, ... Fluctuations in these labels are, in principle, observed as localized sources of radiation.  The expectation is that, even if the macroscopic geometry is symmetric, the quantum geometry will be fluctuating on the microscopic scale" \cite{MS}. Our work here explores the idea of assuming that the Planck scale geometric degrees of freedom are intrinsically quantum and obey spin-determined quantum statistics. The later is admittedly a strong assumption. Without a fuller understanding of the theory's dynamics it is currently not possible to establish a correlation between spin and the statistics of the geometric quanta. Nevertheless we consider it worthwhile to explore variations on the properties of geometric particles. In this paper we find that in the large area limit - the relevant limit to verify the Bekenstein-Hawking entropy - the system condenses to a relatively low entropy state. In the second paper of this series this condensed state provides a physical framework to explore near horizon fluctuations and their contribution to black hole entropy \cite{MRT}.

In the next section we review the context for the statistical models including the relevant kinematic states of LQG and the energy for high acceleration observers. In section \ref{statistics} we calculate the behavior of the ensemble in the high acceleration, large area limit.  This includes discussion on condensation, including a phase with Bose-Einstein condensation.  Since the Barbero-Immirzi parameter acts as an effective temperature we discuss its role in the two types of condensation. In this section we also present the calculations for the fermionic ensemble and entropy of mixing.  We find that the large area limit selects a low entropy condensed state with the occupation essentially all in the lowest ($j=1$) bosonic state. We close with some further discussion on the role of the Barbero-Immirzi parameter and the nature of the condensed state.     

\section{Framework}

We assume that the geometry is described by the quantized geometry of (kinematic) loop quantum gravity. States of geometry are given by $\ket{s} = \ket{ { \sf G} \; \vec{j} \; \vec{{\sf I }} \,}$ for graph ${\sf G}$, spin network link spins $\vec{j}$, and intertwiner labels $\vec{{\sf I }}$ (which may be represented as trivalent decompositions of every intersection of valence four or higher) \cite{LCspinnets}.  For the area ensembles used here we omit the graph and intertwiner labels and simplify the notation to  $\ket{\vec{j}\,}$. We assume boundary conditions, such as in the $SU(2)$ Chern-Simons approach \cite{ENP,ENPP}, that give a degeneracy of $2j+1$ for tiles with spin $j$.  This is in contrast to the work of \cite{GNP,asinetal} which assumes a holographic degeneracy.

The area spectrum of a surface $S$ is \cite{LCgeom,ALgeom} \footnote{There may be vertices that lie in the surface but these are neglected here.} 
\begin{equation}
	\label{Area}
	\hat{A_S} \ket{ \vec{j} \, } = a \ket{ \vec{j} \, } = 8 \pi \gamma \hbar \sum_{i \in \{S \cap {\sf G} \}} \sqrt{j_i \left(j_i+1 \right)} \ket{ \vec{j} \, } ,
\end{equation}
where the set of links of the graph ${\sf G}$ that intersect the surface $S$ are enumerated with $i$. Their spins are denoted $j_i$. The area scale $\ell^2 := \gamma \hbar$ differs from the `fundamental Planck area' $\hbar$ by the Barbero-Immirzi parameter $\gamma \neq 0$, which will be a focus of the LQG aspects of the model \footnote{Classically, the theory is equivalent to general relativity for all values of the parameter $\gamma$ \cite{LQGrev}.  However, in the kinematical state space of LQG the Hilbert spaces $\mathcal H_\gamma$ for different $\gamma$'s are unitarily inequivalent representations \cite{Hgammainequiv}.}.  For much of this paper we work with fixed $\gamma = 0.2734\dots$, which is derived in Appendix \ref{mBI} by setting the entropy of the smallest molecule of geometry with non-vanishing volume equal to the corresponding microscopic Bekenstein-Hawking entropy.

We take an operational approach and work with static observers, or in general dynamic spacetime contexts uniformly accelerating observers, who maintain the magnitude of their proper acceleration, $g = (M/r^2) (1-2M/r)^{-1/2}$. For simplicity we study this class of near-horizon or high-$g$ observers, for whom the quasilocal energy takes on a remarkably simple form \cite{FGP,Eg} \footnote{The corrections in the high-$g$ limit are $\tfrac{\ln(g^2 A)}{4g} + O(1/g)$.}
\begin{equation}
	\label{Eg}
	E_g = \frac{g A}{8 \pi}.
\end{equation}
The energy is proportional to the area. We explore statistical ensembles of area quanta with this energy.  We call these quanta ``tiles." In the literature tiles are sometimes referred to as punctures or plaquettes. With the area spectrum of (\ref{Area}), the Hamiltonian for the model simply inherits its spectrum from the area operator. In the high-$g$ limit $\hat{H} := E_g(g, \hat{A})$ has spectrum
\begin{equation}
	\label{highghamiltonian}
	\hat{H} \ket{\vec{j} \, } =  \sum_{i=1}^N E_{j_i} \ket{\vec{j} \, } = \epsilon \sum_{i=1}^N \sqrt{j_i(j_i+1)} \ket{\vec{j} \, },
\end{equation}
where $\epsilon = g \gamma \hbar$. 

Due to the Hawking-Unruh effect the accelerating observers are bathed in thermal radiation. In the high-$g$ limit the local Hawking temperature and the Unruh temperature are equal. For acceleration $g$ the inverse temperature is
\[
	\beta_U = \frac{2 \pi}{\hbar \, g} \simeq \frac{1}{T_{loc}} . 
\]

Traditionally in LQG black hole entropy calculations the tiles are treated as distinguishable particles (e.g. see \cite{entropycarlo,entropykirill,ABK,MS,alejandro_rev}).  This approach is reviewed in Appendix \ref{distinguish} where we see that the ratio of the single particle partition function $Z_1$ to the number of particles $N$ is small, certainly not satisfying $Z_1/N \gg 1$, which characterizes classical treatments of statistical constituents. Given this we explore an ensemble of indistinguishable quantum constituents in the next section.  

We make the strong assumption that the quanta of geometry obey Fermi or Bose distributions depending on their spin, suggesting that `local Lorentz invariance' (LLI) holds to the quantum gravity scale and the spin statistics theorem holds for the tiles themselves. Since LLI must be addressed with the theory's dynamics, which is currently incomplete, we cannot support this with an argument. However, there are some indications that the theory is compatible with LLI.  Rovelli and Speziale observed that just as the quantization of angular momentum is compatible with rotational covariance, there is in principle no incompatibility between discrete areas and LLI; expectation values  of an operator with a discrete spectrum may be continuous \cite{RS}. More recently within the context of loop-quantized parametrized field theory with matter, Varadarajan has shown that the nonseperability of the Hilbert space can resolve the apparent conflict between discrete area spectra and Lorentz covariance \cite{madhavan}. Additionally, if there was a loss of LLI on the scale of the microscopic constituents we would expect modified dispersion relations.  (In a LQG-like context see for instance \cite{AM-TUphot}.) Current observations do not support evidence of modified dispersions relations. Modified dispersions relation effects originating at the Planck scale both at leading order $p^3/\sqrt{\hbar}$ and at next to leading order $p^4/\hbar$ are constrained by observation \cite{QGPrevcosmo,FHM,SL2013update,HL} and fine-tuning considerations \cite{CPSUV}.

The observers are characterized by their proper acceleration and the spacetime is characterized by the (asymptotic) mass. Since area is directly observationally accessible to the observers, we use area instead of mass. The statistical ensemble is defined by acceleration (or equivalently temperature), area,  Barbero-Immirzi parameter (which sets the microscopic granularity), and the chemical potential.

\section{Quantum statistical mechanics}
\label{statistics}

With the quasilocal Hamiltonian of equation (\ref{highghamiltonian}) and the assumption of spin-linked quantum statistics the partition function $Z={\rm Tr} \left[ e^{-\beta \hat{H}} \right]$ is the product $Z_+ Z_-$ with
\begin{equation}
	\label{PF}
	Z_\pm = \prod_j \left[ 1 \pm \exp(- \beta E_j + \beta \mu_\pm) \right]^{(-1)^{(2j+1)}}
\end{equation}
with $+$ for fermions and $-$ for bosons.  For convenience we use dimensionless chemical potentials $\tilde{\mu}_\pm = \mu_\pm/\epsilon$. The occupation number of tiles in the high-$g$ limit becomes 
\begin{equation}
	\ev{n_j}_\pm = \frac{ 2j+1 }{e^{\beta \epsilon \left(  \sqrt{j(j+1)} - \tilde{\mu}_\pm \right)} + (-1)^{2j+1}}.
\end{equation}

In the usual low temperature limit of statistical mechanics, bosons condense in the lowest possible state and the bosonic chemical potential is approximately equal to the ground state energy while fermions stack up in energy and the fermionic chemical potential is approximately the Fermi energy. Since high-$g$ observers experience a thermal bath at the Unruh temperature, $T_{loc} \propto g$, the situation differs for these observers.  The energy (\ref{Eg}) itself contains a factor of $g$ so the acceleration (or temperature) cancels. The Boltzmann weight no longer has the usual temperature dependence. Instead, the analogous Boltzmann factor is fixed at $e^{-\beta \epsilon} = e^{-2 \pi \gamma} \simeq 0.18$ with the Barbero-Immirzi parameter given by the argument in Appendix \ref{mBI}. The ``effective temperature" is neither ``hot" ($e^{-\beta \epsilon} \rightarrow 1$) nor ``cold" ($e^{-\beta \epsilon} \rightarrow 0$).   

Since the partition function of equation (\ref{PF}) is a product, the grand partition function, and following physical quantities, are sums of bosonic and fermionic contributions. For the rest of this section we discuss bosons (henceforth labeled $b$) and fermions (labeled $f$), before briefly discussing the entropy of mixing.

\subsection{Boson distributions}

To match the regime of validity of the Bekenstein-Hawking entropy we need to find the distributions for large area black holes (in the Planckian sense).  For readers less interested in the details, the result is given in equation (\ref{bentropyfinal}). 

The boson partition function is
\begin{equation}
	\label{bosonic}
	Z_b = \prod_{j=1}^\infty \left[ 1 - e^{-\beta \epsilon \left( \sqrt{j(j+1)} -\tilde{\mu}_b \right) } \right]^{-(2j+1)},
\end{equation}
where the sum is over integer $j$. The grand potential becomes
\[
	\Omega_b = - \frac{1}{\beta} \ln Z_b = \frac{1}{\beta} \sum_{j=1}^\infty (2j+1) \ln \left[ 1 - e^{-\beta \epsilon \left( \sqrt{j(j+1)} -\tilde{\mu}_b \right) } \right].
\]
The average number of particles is
\begin{equation}
	\ev{n}_b = - \left( \frac{ \partial \Omega_b}{\partial \mu_b} \right) = \sum_{j=1}^\infty \frac{2j+1}{e^{\beta \epsilon ( \sqrt{j(j+1)} -\tilde{\mu}_b)} -1 } = \sum_{j=1}^\infty \ev{n_j}_b,
\end{equation}
where the average occupation for spin $j$ is
\begin{equation}
	\label{nave}
	\ev{n_j}_b = \frac{ 2j+1}{e^{\beta \epsilon ( \sqrt{j(j+1)} -\tilde{\mu}_b)} -1}.
\end{equation}
The bosonic area is
\begin{equation}
	\frac{\langle A \rangle_b}{8 \pi \gamma \hbar } =  \sum_{j=1}^\infty \frac{ (2j+1) \sqrt{j(j+1)} }{e^{\beta \epsilon (\sqrt{j(j+1)}- \tilde{\mu}_b)}-1} 
\end{equation}
which, at the Unruh temperature, becomes
\begin{equation}
	\label{area}
	\frac{\langle A \rangle_b}{8 \pi \gamma \hbar} = \sum_{j=1}^\infty \frac{ (2j+1) \sqrt{j(j+1)} }{e^{2 \pi \gamma ( \sqrt{j(j+1)}-\tilde{\mu}_b)}-1}.
\end{equation}
For physical (positive, nonvanishing) areas, the bosonic chemical potential $\tilde{\mu}_b$ has a maximum of $\sqrt{2}$ (and so $\mu_b = \sqrt{2} g \hbar \gamma $). At fixed Barbero-Immirzi parameter, large areas occur when the chemical potential is near this value. As $\tilde{\mu}_b$ approaches $\sqrt{2}$ the particles condense into the $j=1$ state, and the area becomes macroscopic.  Defining $\delta$ to be the difference between this maximum $\tilde{\mu}_b$ and the actual value so that, $\tilde{\mu}_b = \sqrt{2} - \delta$, we can separate the divergent $j=1$ contribution from the rest. In the high-$g$ and large area limit one obtains
\begin{equation}
	 \frac{\langle A \rangle_b}{8 \pi \gamma \hbar} = \sqrt{2} \ev{n_1}_b + \sum_{j=2}^\infty \sqrt{j(j+1)} \ev{n_j}_b \simeq \frac{3 \sqrt{2}}{2 \pi \delta} + O(1) .
\end{equation}
Due to the area scale involved -- for example a solar mass black hole has ratio $A_{\odot}/ \hbar \sim 10^{77}$ -- the order one remainder can been neglected \footnote{The integral approximation for the correction is 
\[
	\frac{1}{4 (\pi \gamma)^3} \int_{\alpha(\sqrt{3} -1)}^\infty \frac{(x + \alpha)^2}{e^x-1} dx \simeq 2.02,
\]
where $\alpha = 2\sqrt{2} \pi \gamma$.}. Clearly, the relevant large area limit is $\delta \rightarrow 0$ and $\delta$ (and $\mu_b$) differ from among black holes of differing area.  

The total number of bosonic tiles is 
\begin{equation}
	\label{Nb}
	\ev{N}_b = \sum_{j=1}^\infty \frac{2j+1}{e^{2 \pi \gamma \sqrt{j(j+1)} - \tilde{\mu}_b)}-1} \simeq \frac{1}{2 (\pi \gamma)^2} \int_{2 \pi \gamma \delta}^\infty \frac{x+2 \pi \gamma \delta}{e^x-1} dx \simeq \frac{3}{2 \pi \gamma \delta}
\end{equation}
in the large area limit.

Similarly separating out the $j=1$ average occupation number from the rest gives
\begin{equation}
	\ev{n_1}_{b} \simeq \frac{3}{2 \pi \gamma \delta} .
\end{equation}
The sum of all higher level occupation numbers is
\begin{equation}
	 \ev{n_{j>1}}_b = \sum_{j=2}^\infty \ev{n}_j \simeq \frac{1}{2 (\pi \gamma)^2} \int_{\alpha(\sqrt{3} -1)}^\infty \frac{(x + \alpha)}{e^x-1} dx \simeq 0.96,
\end{equation}
where $\alpha = 2\sqrt{2} \pi \gamma$, in the large area limit. It is clear that in this limit the $j=1$ level is occupied while the higher levels are relatively unoccupied; $\ev{n_{j>1}}/\ev{n}_b \simeq \ell^2/\ev{A}_b \rightarrow 0$.

The bosonic entropy is 
\begin{equation}
	\label{bentropy}
	S_b = \beta^2 \frac{ \partial \Omega_b}{\partial \beta} = \beta \ev{U}_b + \ln Z_b - \beta \mu_b \ev{N}_b.
\end{equation}
The energy of the system is
\[
	\ev{U} = \epsilon \sum_{j=1}^\infty  \sqrt{j(j+1)} \ev{n_j} = \frac{g \ev{A}}{8 \pi}
\]
so at the Unruh temperature the term in the entropy, $\beta_U \ev{U}_b = \ev{A}_b/4 \hbar$, is exactly the Bekenstein-Hawking entropy without fixing the Barbero-Immirizi parameter.  This is a direct consequence of the form of the energy $E_g$ in this limit.  The relation holds for both bosonic and fermionic contributions.  In the calculation for distinguishable particles in Appendix \ref{distinguish} this relation drives the $S \propto A$ relation. Here the additional terms in the entropy cancel this contribution leading to anemic growth.  The last term at the Unruh temperature is 
\[
	- \beta_U \mu_b  \ev{N}_b \simeq   - \frac{ 3 \tilde{\mu}_b }{\delta},
 \]
which in the large area limit becomes $- \ev{A}_b / 4\hbar$, exactly canceling the Bekenstein-Hawking result from the first term. Physically this is expected in that the system has condensed into the low entropy $j=1$ state.

\begin{figure}
	\begin{tabular}{cc}
	\includegraphics[scale=.6]{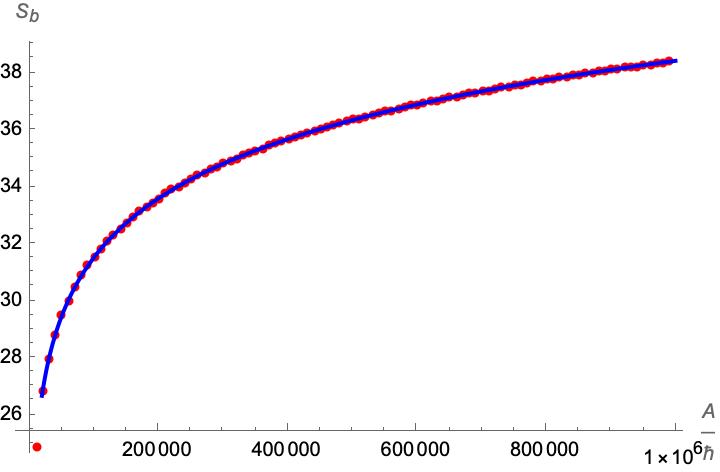} & \includegraphics[scale=.6]{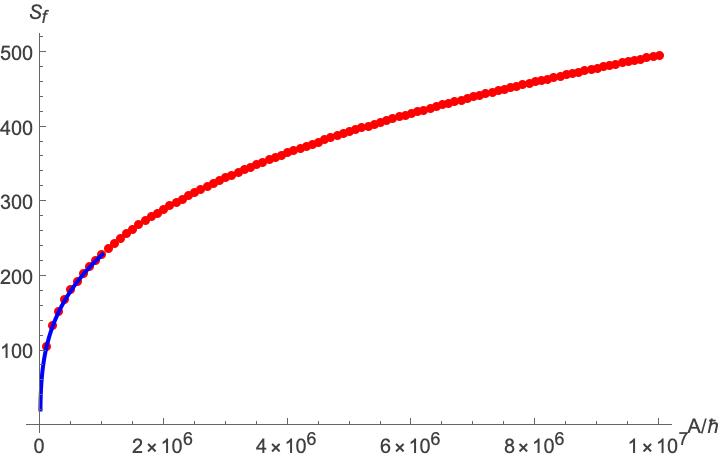} \\
	(a) & (b) 
	\end{tabular}
	\caption{\label{intapprox} Entropy in the high-$g$, large area limit as a function of area.  In (a) the numerically computed bosonic entropy $S_b$ is shown with the red filled circles. The integral approximation of the bosonic entropy,  $3 \ln (A/3a_1)+7$, is shown with the blue solid line. Similarly in (b) the numerically computed fermionic entropy $S_f$ is shown with the integral approximation with the same choice of filled dots for the numerical approximation and the solid line for the integral approximation. }
\end{figure}

The log term
\[
	\ln Z_b = - \sum_{j=1}^\infty (2j+1) \ln \left( 1 - e^{-2 \pi \gamma \left( \sqrt{j(j+1)} - \tilde{\mu}_b \right)} \right),
\]
gives the subleading correction. Splitting off the log divergent term and using the integral approximation gives 
\begin{equation}
	\begin{split}
	\ln Z_b & \simeq - 3 \ln ( 2 \pi \delta ) - (\sqrt{2} \pi \gamma)^{-2} \int_{2 \sqrt{2} \pi \gamma (\sqrt{3} - 1) }^{\infty} (x + 2 \sqrt{2} \pi \gamma) \ln (1 - e^{-x}) dx \\
	& \simeq 3 \ln \left( \frac{ \ev{A}_b}{ 24\sqrt{2} \pi \gamma \hbar} \right)  
 	\end{split}
\end{equation}
in the large area limit. In sum the first and last terms of the entropy in equation (\ref{bentropy}) cancel and the remaining log term gives the total bosonic entropy
\begin{equation}
	\label{bentropyfinal}
	S_b  \simeq 3 \ln \left( \frac{ \langle A \rangle_b}{3 a_1} \right)
\end{equation}
in the high-$g$, large area limit.  We have used the condensed state tile area $a_1 = 8 \sqrt{2} \pi \gamma \hbar$. Figure \ref{intapprox}(a) shows that the integral approximation for the entropy holds even for relatively small black holes.

Notice that since the energy scales with $g$ - effectively the temperature, $\beta E$ becomes a fixed number and ``temperature is irrelevant".This is in contrast with BE condensation where the expression retains a temperature dependence. 

We close this section with remarks on the role of the Barbero-Immirzi parameter. For comparison to the usual case of a Bose gas it is useful to let $t=1/2\pi \gamma$. This ``temperature" is only analogous to a physical temperature in the role it plays in the distribution. Bose-Einstein condensation occurs in an analogous way to usual Bose systems when one determines the critical temperature by fixing the total number of particles.  Here we fix the observable area $A$ rather than the number of tiles but the approach is similar. 
For a fixed, large area $A$ the Barbero-Immirzi parameter may be determined as follows. The ratio of area to Planck area is given by
\begin{equation}
	\frac{A}{\hbar} =\frac{2}{(\pi \gamma)^2} \int_{2 \sqrt{2} \pi \gamma}^\infty \frac{x^2}{e^x-1} dx,
\end{equation}
where we have used $\ev{A}_b$ of equation (\ref{area}) and the integral approximation at vanishing chemical potential. The integral over $x=2 \pi \gamma j$ evaluates to $2 \zeta(3)$ yielding a critical Barbero-Immirzi parameter of $\gamma_c =(2 \sqrt{\zeta(3)}) / \pi (\hbar/A)^{1/2} \simeq 0.7 (\hbar/A)^{1/2}$. Clearly, $\gamma_c$ is vanishingly small for all but microscopic black holes ($\gamma_c \sim 10^{-77}$ for solar mass black holes). Since the Barbero-Immirzi parameter is thought to be a fundamental parameter of the theory rather than a macroscopic quantity associated to specific systems, the physical role of such a critical parameter is unclear.  Nonetheless the analogous Bose-Einstein condensation behavior occurs: The system condenses to the ground ($j=1$) state at a finite ``temperature" $t_c=1/2 \pi \gamma_c$ and the occupation number $\ev{n_1}_b$ increases as $(\gamma_c/\gamma)^2$ below the critical temperature.  There is a notable difference: Due to the scaling of area with $\gamma$, the occupation number $\ev{n_1}$ as a function of $\gamma$ does not monotonically increase as $t$ is lowered but instead achieves a maximum $\ev{n_1}|_{\gamma_*} \simeq 0.16 \sqrt{A/\hbar}$ at $\gamma_*=\sqrt{3} \, \gamma_c$. This critical behavior could become physically relevant if there was a fundamental lower limit on the mass of black holes. The Barbero-Immirzi parameter used in most of this work, $\gamma \simeq 0.2734$, corresponds to a critical Barbero-Immirzi parameter for a roughly one Planck mass black hole.  See the section \ref{discussion} for more discussion on this perspective.

There is another way to achieve large areas, sending $\gamma \rightarrow 0$ and $j \rightarrow \infty$ while keeping $\gamma j$ fixed, which is often used in the semiclassical limit in spin foams.  Taking $\mu_b=0$  in this limit one finds $\ev{A}/\hbar  \sim \gamma^{-2} \sim N$ and the entropy agrees with the Bekenstein-Hawking result without further fixing Barbero-Immirzi parameter.  For a solar mass black hole in this double-scaling limit $\gamma \sim 10^{-39}$ and the spin would be $j \sim 10^{37}$ for a Planck scale tile $a_j \sim \hbar$.  Naively the curvature, measured for example by a Wilson loop around the large spin link, appears far from zero and thus differs from the expected relatively smooth geometry of the semiclassical limit. Other studies also suggest that the semiclassical limit of LQG is found in highly flocculent networks with low spins. This arises due to the `minimum angle-volume bound';  in reducing the granularity of the spectrum of the angle operator \cite{angle} one necessarily increases the volume of the atom of geometry \cite{MS}. 

Finally, due to the small value of $\delta$ in the large area limit it is worth exploring the role of the Barbero-Immirzi parameter in the integral approximations. At maximum chemical potential $\tilde{\mu}_b=\sqrt{2}$ the total number of tiles and area satisfy
\begin{equation}
	N \simeq \frac{I_N(\gamma)}{2\pi^2 \gamma^2} \quad \text{and} \quad A\simeq\frac{2\hbar}{\pi^2\gamma^2}I_A(\gamma)
\end{equation}
where 
\[
	I_N(\gamma) := \int_{0}^\infty \frac{x+2\pi\gamma\sqrt{2}}{e^x-1} dx \quad \text{and} \quad I_A(\gamma) := \int_{0}^\infty \frac{(x+2\pi\gamma\sqrt{2})^2}{e^x-1} dx.
\]
The divergence of these integrals is exactly what we expect when setting $\Tilde{\mu}_b=\sqrt{2}$. With this in mind, and noting that 
\begin{equation}
	\frac{d I_A}{d \gamma}=4\pi \sqrt{2}I_N(\gamma)
\end{equation}
one can obtain 
\begin{equation}
	N =  \left( \frac{1}{a_1} \right) \left( 1 + \frac{\gamma}{2} \frac{d}{d \gamma} \right) A.
\end{equation}
This relation contains our expected result for condensation, $N=\frac{A}{a_1}$, along with an additional term describing the flow in $\gamma$ of the area. A similar relation holds for energy.

\subsection{Fermionic distribution}

The fermionic case follows a similar method, although the large area limit requires large chemical potential.  It turns out that the chemical potential for large fermionic areas is larger than the bosonic case so fermionic configurations are disfavored and the fermionic contribution will be negligible.  Nonetheless we include the discussion so we can both see the scaling of area and number and use the results in the next section on entropy of mixing.

The fermionic partition function is  
\begin{equation}
	\label{fermionic}
	Z_f = \prod_{j=1/2}^\infty \left( 1 + e^{\beta( \epsilon \sqrt{j(j+1)} -\mu_f)} \right)^{2j+1},
\end{equation}
where is the sum is over half odd integers. The area is 
\begin{equation}
	\label{farea}
	\frac{\ev{A}_f}{ 8 \pi \gamma \hbar} =  \sum_{j=1/2}^\infty \sqrt{j(j+1)} \ev{n_j}_f  =  \sum_{j=1/2}^\infty \frac{\sqrt{j(j+1)} \, (2j+1)}{e^{2 \pi \gamma(\sqrt{j(j+1)}- \tilde{\mu}_f)} +1},
\end{equation}
where the second equality holds at the Unruh temperature. As before, we define a dimensionless chemical potential $\tilde{\mu}_f = \mu_f/\epsilon$ for the fermions. Since the Boltzmann weight is at minimum $e^{\sqrt{3} \pi \gamma} \sim 4$ it is clear that large areas exists only for very large $\tilde{\mu}_f$.  In this case contributions to the sum will extend to large $j$ and the area grows as $j_{max}^3 \sim \tilde{\mu}_f^3$. In fact, 
\begin{equation}
	\ev{A}_f \simeq - \frac{4 \hbar}{(\pi \gamma)^2} \text{Li}_3 (-e^{2 \pi \gamma \tilde{\mu}_f}) \simeq \frac{16 \pi \gamma \hbar}{3} \tilde{\mu}_f^{3} + \frac{4 \pi \hbar}{3 \gamma} \tilde{\mu}_f,
\end{equation}
where $\text{Li}_n (z)$ is the polylogarithm function. 

The total number of fermionic tiles is 
\begin{equation}
	\label{Nf}
	\begin{split}
	\ev{N}_f &= \sum_{j=1/2}^\infty \frac{2j+1}{e^{2 \pi \gamma \sqrt{j(j+1)} - \tilde{\mu}_f)}+1} \simeq \frac{1}{2 (\pi \gamma)^2} \int_{2 \pi \gamma \tilde{\mu}_f}^\infty \frac{x+2 \pi \gamma \tilde{\mu}_f}{e^x+1} dx = - \frac{1}{2 (\pi \gamma)^2} \text{Li}_2 \left( -e^{2 \pi \gamma \tilde{\mu}_f} \right) \\
	& \simeq \tilde{\mu}_f^2 + \frac{1}{12 \gamma^2}.
	\end{split}
\end{equation}
For large area $\ev{N}_f  \simeq \tilde{\mu}_f^2 + (12 \gamma^2)^{-1}$. As we will see soon, it is this fractional scaling with area $\ev{N}_f \sim \ev{A}_f^{3/2}$ that disrupts the $S \propto N$  scaling in the entropy of mixing. 
 
The fermionic entropy is 
\begin{equation}
	S_f = \beta^2 \frac{ \partial \Omega_f}{\partial \beta}  = \beta \ev{U}_f + \ln Z_f - \beta \mu_f \ev{N}_f = \frac{\ev{A}_f}{4 \hbar}  + \ln Z_f(\beta_U) - \beta_U \mu_f \ev{N}_f .
\end{equation}
at the Unruh temperature. As expected, the first term is in the Bekenstein-Hawking form.  At the observers' Unruh temperature the third term simplifies to 
\begin{equation}
	 - \beta \mu_f \ev{N}_f \simeq  - \frac{3 \ev{A}_f}{8 \hbar},
\end{equation}
which contributes to the cancelation of the Bekenstein-Hawking entropy. The second term 
\begin{equation}
	\ln Z_f(\beta_U) = \sum_{j=1/2}^\infty (2j+1) \ln \left[ 1 + e^{2\pi \gamma \left( \sqrt{j(j+1)} - \tilde{\mu}_f \right) } \right]  
\end{equation}
with the integral approximation becomes
\begin{equation}
	\begin{split}
	\ln Z_f &\simeq \frac{1}{2(\pi \gamma)^2} \int_{-2 \pi \gamma \tilde{\mu}_f}^\infty ( x + 2 \pi \gamma \tilde{\mu}_f ) \ln \left( 1 +e^{-x} \right)  dx = - \frac{1}{2(\pi \gamma)^2} \text{Li}_3(-e^{2 \pi \gamma \tilde{\mu}_f}) \\
	&\simeq \frac{ \ev{A}_f}{ 8 \hbar} + \frac{\pi}{6 \gamma} \tilde{\mu}_f .
	\end{split}
\end{equation}
Thus the leading order $\tilde{\mu}_f^3$ terms, which are proportional to area, cancel leaving the subleading term
\begin{equation}
	S_f \simeq \left( \frac{ \pi^2 \ev{A}_f}{144 \gamma^4 \hbar} \right)^{1/3}
\end{equation}
in the fermionic contribution to the entropy in the large area limit. 

In the large area limit of the bose and fermi ensembles we see that the ratio of bose to fermi chemical potentials is
\begin{equation}
	\frac{\mu_b}{\mu_f} = \frac{ \sqrt{2} \gamma \hbar}{\left(3 \gamma^2 \hbar^2 M^2 \right)^{1/3}} \sim \left( \frac{ \ell^2 }{A} \right)^{1/3}.
\end{equation}
As area grows it is increasingly less energetically favorable to add fermionic tiles. 

Since the ensemble may be a mixture of difference species we need to find the entropy of mixing, which can be large for a relatively small number of one type of particle mixed with a large number of another type.
 
\subsection{Entropy of mixing}

While one could model tile mixtures in a variety of ways, for instance considering tiles differing by spin as distinguishable species, a simple two species will serve to show the behavior in the large area limit.  We model an ensemble of two species of particles, fermions and bosons.  Defining the fraction $x_i = \ev{N}_i/\ev{N}$ and the boson to fermion fraction as $X = \ev{N}_b / \ev{N}_f$, the  entropy of mixing becomes
\begin{equation}
	S_{mix} = - N \left( x_f \ln x_f + x_b \ln x_b \right) = \ev{N}_f \ln \left( 1 + X \right) + \ev{N}_b \ln \left(1 +\frac{1}{X} \right) 
\end{equation}
From equations (\ref{Nb}) and (\ref{Nf})
\[
	X \simeq \frac{3}{2 \pi \gamma \delta \tilde{\mu}_f^2} 
\]
in leading order. So a possible double scaling limit of chemical potentials presents itself.  However, in the large area limit, the system is essentially all in the $j=1$ state; $N\sim \ev{N}_b$ and $X$ diverges.  We recover the $S \sim \ln (A/\ell^2)$ scaling of equation (\ref{bentropyfinal}). Balancing boson and fermion contributions is spoiled by the mismatch between the $\ev{N}_f \sim \ev{A}_f^{2/3}$ scaling for fermions and the $\ev{N}_b \sim \ev{A}_b$ scaling for bosons. 

\section{Discussion}
\label{discussion}

The statistical model of black hole geometry for uniformly accelerating observers presented in this paper is built on the the area quanta of LQG, the quasilocal Hamiltonian of equation (\ref{highghamiltonian}), the spin-determined particle statistics, and the spin projection degeneracy. Notably the physical temperature $\propto g$ cancels in the distribution, as it must: The results hold for all observers in the high-$g$ limit.  In the large area limit the bosonic statistical mechanics yields a low entropic condensed state in which essentially all the tiles are in the lowest, $j=1$, state with area $a_1 = 8 \sqrt{2} \pi \gamma \hbar$, which sets the scale for the microscopic geometry.  The entropy is given in equation \ref{bentropyfinal}. This ``observer condensate" is distinct from the usual Bose-Einstein condensate in that it does not depend on physical temperature. 

In the fermionic case the distribution requires a large chemical potential for large area and any significant occupation in fermionic states is disfavored in the large area limit.  Although entropy of mixing can in principle affect the counting significantly, in the framework of large areas and the $O(1)$ Barbero-Immirzi parameter studied here, entropy of mixing does not contribute significantly to the total entropy. In sum the model of a gas of noninteracting, indistinguishable geometric particles at high-$g$ and large area is a condensate of predominantly $j=1$ tiles. Loosening the restriction on the value of the Barbero-Immirzi parameter yields Bose-Einstein condensate-like behavior with a ``temperature" of $t=1/2\pi \gamma$. We leave the interpretation of this phase to further work. 

 The results of this paper and those of \cite{GNP,asinetal} demonstrate that if one includes fully quantum tiles in the statistical models one must locate the large Bekenstein-Hawking entropy either in huge particle degeneracies such as holographic degeneracies of $\exp(A/4 \hbar)$ or in physical processes elsewhere. In Ref. \cite{MRT} we use this low entropy bose condensate as a starting point to model the fluctuations in the near horizon geometry.

There is a potentially intriguing connection of this condensed $j=1$ state to earlier work. In \cite{olaf} Dreyer noticed that the correspondence principle suggested that the high frequency of quasinormal modes should equal the frequency of the fundamental quantum transition, the exchange of the $j=1$ tile with the exterior.  Following early numerical work by Nollert \cite{nollert}, Hod noticed that the real part of the large overtone frequencies of quasinormal modes was given by $\ln 3/8 \pi M$ \cite{hod}. The tile exchange idea can be used to fix the Barbero-Immirzi in the uniformly accelerating observer's frame as follows. For the high-$g$ observers the real part of the asymptotic (in overtone) quasinormal mode spectrum is
\begin{equation}
	\omega_{QNM} =  \frac{\ln 3}{8 \pi M} \left(1 - \frac{2M}{r} \right)^{-1/2} \simeq \frac{ g \ln 3}{2 \pi}.
\end{equation}
Dreyer suggested that this classical frequency should correspond to the quantum transition of a tile of minimum $j$.  Using equation (\ref{Eg}) and $j=1$ we see that 
\begin{equation}
	\Delta E_g = \frac{g \Delta A}{8 \pi} = \frac{g a_1}{8 \pi} = g \sqrt{2} \gamma \hbar = \hbar \omega_{a_1}
\end{equation}
Equating the two frequencies, $\omega_{QNM} = \omega_{a_1}$ gives  $\gamma = \ln(3)/2 \pi \sqrt{2} \simeq 0.12$. The condensed state of $j=1$ tiles naturally provides a physical mechanism for a minimum spin of 1, rather than a switch in the underlying internal symmetry from $SU(2)$ to $SO(3)$ suggested in Dreyer's work.  In Ref. \cite{MRT} we propose that the degrees of freedom or ``internal configurations" are neither exchanges of microscopic constituents nor degrees of freedom inside the horizon but rather are near-horizon fluctuations.

\begin{acknowledgments}
It is a pleasure to thank Aben Carrington, Steven Cunden, Christopher Lockwood, John Pikus, and Andrew Projansky for previous work and fruitful discussions.  SM thanks Alejandro Perez for a discussion years ago that led to this work and to Madhavan Varadarajan for helpful discussions on Lorentz invariance, observers and the PPT model.  We thank Hamilton College for generous financial support.
\end{acknowledgments}

\appendix

\section{Fixing the Barbero-Immirzi parameter microscopically}
\label{mBI}

The Barbero-Immirzi parameter $\gamma$ may be fixed by the following argument.  Suppose that the Bekenstein-Hawking proportionality $S=A/(4\hbar)$ holds down to the smallest molecule of space that has nonvanishing area, angle and volume. This is a quantum tetrahedron with face areas $a_{1/2} = 4 \sqrt{3} \pi \gamma \hbar$ for spin$-1/2$. Although in the bulk the magnetic quantum numbers are pure gauge, we assume boundary conditions on the surface of the quantum tetrahedron such that the area density, $n_a E^{ai}$, becomes an boundary observable and the magnetic quantum numbers must be specified to completely determine the state \cite{samdis,HMboundary}.  In this case each face has degeneracy of 2. The number of microstates of the tetrahedron is then $\Omega = 2^4 4!$ \cite{simonedis}. Thus,
\begin{equation}
	\label{gammavalue}
	S = \ln \Omega = 7 \ln 2 + \ln 3 = \frac{A}{4 \hbar} = 4 \sqrt{3} \pi \gamma \implies \gamma = \frac{7 \ln 2 + \ln 3}{4 \sqrt{3} \pi} \simeq 0.2734\dots
\end{equation}
This value differs slightly from the Barbero-Immirzi parameter derived through state counting and matching the Bekenstein-Hawking entropy:  In Refs \cite{DL,M} the authors find $\gamma =0.2375...$ which differs from this by a relative error of about 13\%. Using Chern-Simons theory researchers find $\gamma=0.2741...$, which differs from this by a relative error of about 0.2\%.\footnote{See page 28 and footnote 15 in the review \cite{alejandro_rev}} Of course, one might expect that the macroscopic value of Newton's constant would be renormalized relative to the microscopic constant effectively used here.

\section{The entropy for distinguishable tiles}
\label{distinguish}
The partition function for a distinguishable, noninteracting gas of particles with energy $E_g = E_j$ is
\begin{equation}
	\label{Zdist}
	Z_d = Z_1^N,
\end{equation}
where the single particle partition function is
\begin{equation}
	\label{Z1}
 	Z_1(\beta) = \sum_{j=1/2}^\infty (2j+1) e^{-\beta E_j}.
\end{equation}
The sum is over spins $j = \tfrac{1}{2} ,1, \tfrac{3}{2}, ...$ .  The single particle entropy is 
\begin{equation}
	\label{entropy_1p}
	S_1 = \beta \ev{U}_1 + \ln Z_1 = \frac{\beta \epsilon}{8 \pi \gamma \hbar} \ev{A}_1 + \ln Z_1.
\end{equation}
The ensemble entropy at the Unruh temperature is 
\begin{equation}
	\label{entropy_dist}
	S_d = \beta_U \ev{U}_1 + \ln Z_1(\beta_U)  = \frac{\ev{A}}{4 \hbar}  + N\ln Z_1 (\beta_U).
\end{equation}
If the log of the single particle partition function vanishes then $S_d$ is equal to the Bekenstein-Hawking entropy.  This can be done by fixing the Barbero-Immirzi parameter via
\begin{equation}
	\label{Z1exact}
		Z_1 (\beta_U) = \sum_{j=1/2}^{\infty} (2j+1) e^{-2\pi \gamma \sqrt{j(j+1)}} =1.
\end{equation} 
The result  is approximately $\gamma \simeq 0.2741$ and may be obtained numerically by truncating the above sum and checking that the value of $\gamma$ to the desired precision does not depend on the cutoff. This value differs from the microscopic value or equation (\ref{gammavalue}) by a relative error of about 0.2\%. 

The Domagala and Lewandowski (DL) case of Ref. \cite{DL} (without the projection constraint) can be constructed similarly.  In the DL model states are all {\em signed} values of $m_j$.  Effectively this means that the partition function each ``atom" is two copies of the single partition function without the $2j+1$ degeneracy since this is now redundant. Using the same energy of equation (\ref{Eg}) and letting the partition function for positive $m_j$ be 
\begin{equation}
	\label{Z1p}
 	Z_{o}(\beta) = \sum_{m_j=1/2}^\infty e^{-\beta g A(m_j)/ 8 \pi}
\end{equation}
with $A(m_j) = 8 \pi \gamma \hbar \sqrt{|m_j|(|m_j|+1)}$, the single particle partition function is the $Z_{1DL} = Z_o^2$.  The DL model partition function is 
\begin{equation}
	\label{ZDL}
	Z_{DL} = Z_{1DL}^N.
\end{equation}
The calculation of the entropy is as above but now the condition on the log is 
\begin{equation}
		Z_{DL1} (\beta_U) = 2 \sum_{m_j=1/2}^{\infty} e^{-2\pi \gamma \sqrt{m_j(m_j+1)}} =1
\end{equation} 
with the result that the Barbero-Immirzi parameter is $\gamma_{DL} \simeq 0.2375$ as in \cite{DL,M}.

\end{document}